\documentclass[%
aip,
amsmath,amssymb,
reprint
]{revtex4-1}
\usepackage{graphicx}% Include figure files
\usepackage{dcolumn}% Align table columns on decimal point
\usepackage{bm}% bold math
\usepackage{amsmath}
\usepackage{color}
\usepackage[normalem]{ulem}
\usepackage{balance}
\usepackage[T1]{fontenc}

\newcommand{\mycite}[1]{\scalebox{1.3}[1.3]{\raisebox{-0.80ex}{\cite{#1}}}}
\begin{document}

\preprint{AIP/123-QED}

\title{Unified Treatment for Scattering and Photoluminescence Properties of Strongly Coupled Metallic Nanoparticle Chains based on a Coupling Classic Harmonic Oscillator Model}% Force line breaks with \\

\author{Yuqing Cheng}\affiliation{School of Mathematics and Physics, University of Science and Technology Beijing, Beijing 100083, China}%
\author{Mengtao Sun}\thanks{mengtaosun@ustb.edu.cn}
\affiliation{School of Mathematics and Physics, University of Science and Technology Beijing, Beijing 100083, China}%
\affiliation{Collaborative Innovation Center of Light Manipulations and Applications, Shandong Normal University, Jinan, 250358, China}


\begin{abstract}
We present a multimer coupling classic harmonic oscillator model to reveal the scattering and photoluminescence (PL) properties of metallic nanoparticle chains. Taking particle number from 1 to 6 as examples, we compare the calculated spectra with the experimental ones from other researchers' work, and they agree well with each other. Furthermore, scattering and PL properties are analyzed carefully varying with particle number $n$, coupling strength $g$ and effective free electron number $N$. Results indicates larger red-shift and smaller full width at half maximum (FWHM) of the scattering spectra with larger $n$ or/and larger $g$. Meanwhile, the splitting of PL modes increases as $g$ increases, and the amplitudes are dependent on the excitation wavelength.
This classic model is simple and shows a unified treatment for understanding the scattering and PL properties of multimer coupled systems.
\end{abstract}

\maketitle

%\tableofcontents

%\section{\label{sec:Introduction}Introduction}
%\textbf{\textit{Introduction}}.---
Growing interest in coupled systems, especially strongly coupled metallic nanostructures, has been realized by numerous investigators, because the ability and potential of them in optical devices makes it possible to tune the electric field at subwavelength scale. Particularly, localized surface plasmon resonance (LSPR) of metallic nanostructures, e.g., gold nanorods and nanospheres etc., are widely investigated. Due to the amazing optical properties of LSPR, e.g., strongly enhancement and highly confinement of electric field, it has been applied in numerous technological applications such as biosensing \cite{appbio1,appbio2,appbio3,appbio4}, optical recording \cite{appor1,appor2}, optical waveguiding \cite{appowg1,appowg2,appowg3}, nonlinear optics \cite{appnlo1,appnlo2,appnlo3}, and nano-optical devices \cite{appnod1,appnod2,appnod3}.

Multiple nanoparticles including dimers result in new LSPR properties when they are strongly coupled with each other, compared with individual nanoparticle. This phenomenon has been widely investigated both experimentally and theoretically by researchers. S. Biswas $et\ al$. employ self-assembled gold nanorod heterodimers to achieve plasmon-induced optical transparency (PIT) uniquely in the visible wavelengths with a slowdown factor of 10 and extreme dispersion \cite{couple1}. L.-J. Black $et\ al$. tune the polarization conversion of coupled gold nanorod dimer by controlling the gap width. They conclude that much higher conversion efficiencies may be obtained in ``kissing'' antennas \cite{couple2}. C.-Y. Tsai $et\ al$. investigate the optical properties of gold nanoring dimers and find that LSPR peak of them strongly depends on the polarization direction and gap distance, i.e., longitudinal and transverse polarizations result in the red-shift and blue-shift, respectively, when the nanorings approach each other \cite{couple3}. P. Mulvaney $et\ al$. investigate the scattering properties of linear chains of strongly coupled gold nanoparticles with particle number from 1 to 6 by employing self-assembly techniques. They obtain the red-shift of LSPR peak when particle number increases and conclude that a maximum resonance wavelength should be at a chain length of 10--12 particles \cite{couple4}. A. M. Soehartono $et\ al$. employ tall nanosquare dimer for biosensing applications, where the resonance mode and hybrid modes are investigated and the longitudinal mode is found to be red-shifts when increasing the height \cite{couple5}. L. S. Slaughter $et\ al$. investigate LSPR properties of symmetry breaking gold nanorod dimers with different configurations. They find that symmetry breaking has the strongest effect on the collective plasmon modes \cite{couple6}. S. Yoo $et\ al$. report the novel strategy for the synthesis of complex 3-dimensional (3D) nanostructures and apply them to surface enhanced Raman scattering (SERS). The coupling phenomenon is illustrated to be directly related to the intrananogap and interior volume size \cite{couple7}.
All these obtained coupling phenomena have pointed to the fact that strong coupling indicates large splitting of LSPR peak. Explanations based on simulations and models have been done by these researchers. Particularly, the the models usually provides coupled LSPR peak positions. However, the spectra shapes of these coupled systems are seldom estimated in theory. In a previous work, we investigate the spectra of metallic nanoparticles with individual mode without any coupling \cite{PL0}. In another previous work, we only consider the dimer case for the coupled system. Multi-particle coupled system was not realized due to its complexity \cite{PL01}.

In this study, we investigate a metallic multi-particle system which is arranged as a linear chain. We develop a coupling classic harmonic oscillator model for multi-particle to study the optical properties. Taking 1 to 6 particles as examples, we carefully analyze the scattering and PL properties varying with particle number $n$, coupling strength $g$ and effective free electron number $N$. The theoretical results illustrate that, for scattering spectra, LSPR peak red-shifts and full width at half maximum (FWHM) decreases as $n$ or/and $g$ increases; for PL spectra, the splitting of the modes gets larger with larger $g$, and the amplitudes are dependent on the excitation wavelength. Comparisons between our model and experimental data from Mulvaney's work \cite{couple4} demonstrate that our model is practical and accuracy. This
work would give rise to understanding complex coupling phenomena more deeply.

~\\ \indent
%\section{\label{sec:Model}Model}
%\textbf{\textit{Model}}.---
An individual metallic nanostructure with single mode can be treated as an oscillator with its intrinsic frequency $\omega_0$ and damping coefficient $\beta_0$, the behavior of which has been discussed in our previous work \cite{PL0}. When it comes to two coupled nanostructures (assume these two are the same), the coupling coefficients between them can be evaluated as \cite{PL01}:
\begin{equation}
g^2=\frac{1}{\kappa}(\frac{c}{r_0})^3,\ \ \gamma=\frac{1}{\kappa}(\frac{c}{r_0})^2,
\label{eq:g}
\end{equation}
where $\frac{1}{\kappa}=\frac{N e^2}{2\pi \varepsilon_0 m_e c^3}$.
Here $N$ is the effective number of free electrons of each oscillator, $m_e$ is the mass of electron, $r_0$ is the distance between the coupled oscillators, $\varepsilon_0$ is the permittivity of vacuum, and $c$ is the velocity of light in vacuum.
\begin{figure}[tb]
\includegraphics[width=0.4\textwidth]{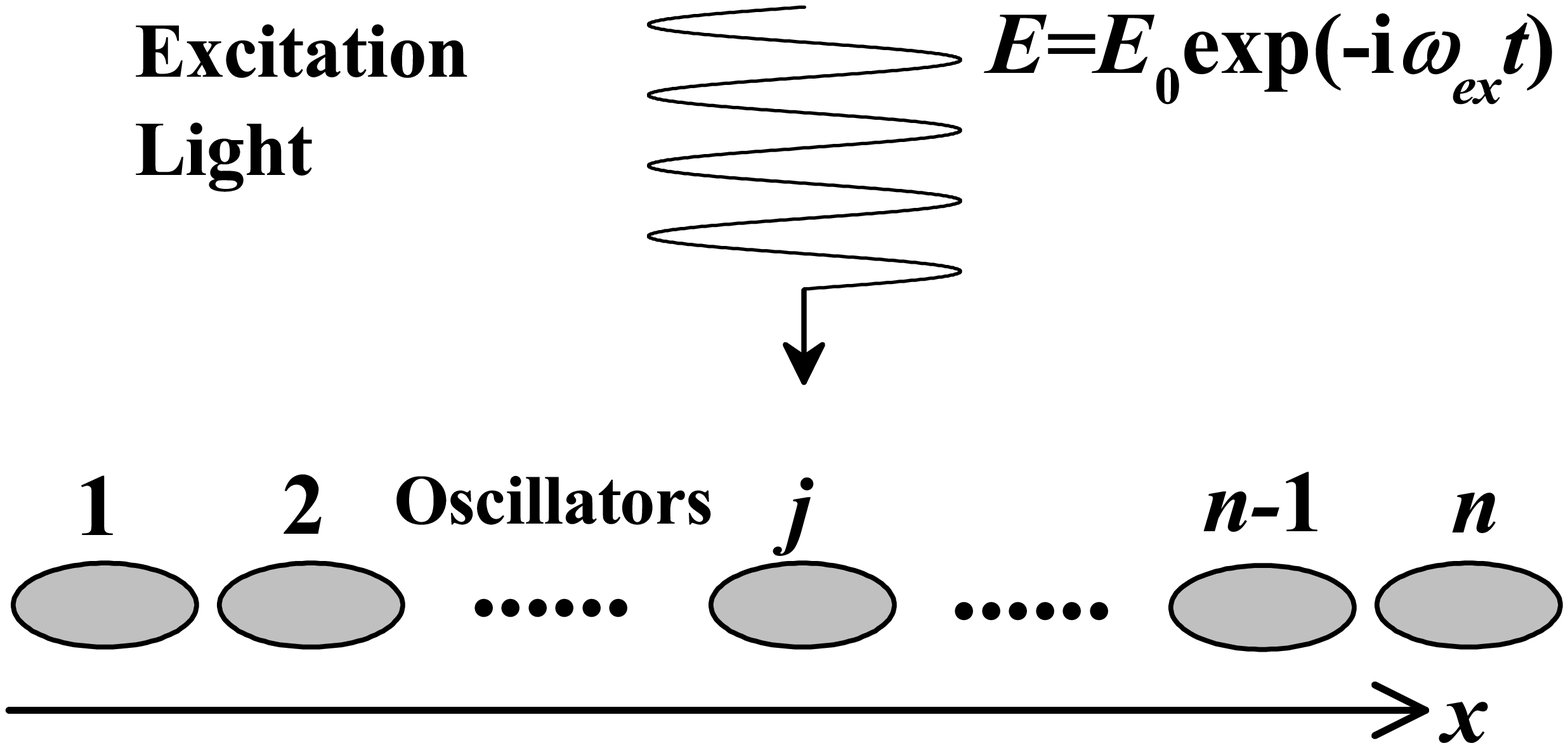}
\caption{\label{fig:Schematic} Schematic of the coupling classic harmonic oscillator model, which is constituted by $n$ oscillators arranged along $x$-axis. Each ellipse stands for an individual oscillator that oscillates along $x$-axis. The $x$-polarized excitation light with amplitude $E_0$ and frequency $\omega_{ex}$ illuminates the system.
}
\end{figure}
Now we consider a metallic nanostructure chain consisted by $n$ nanoparticles arranged in a line. The schematic is shown in Fig. \ref{fig:Schematic}. The $x$-polarized incident light excites the system with frequency $\omega_{ex}$ and amplitude $E_0$. Due to the fact that the coupling strength $g$ decreases rapidly with the increase of $r_0$, we only consider the interaction between the neighboring particles. For the simplest case, we assume that all the particles are the same. Therefore, the dynamical equations of these oscillators should be in this form:
\begin{equation}
\begin{aligned}
&\ddot{x}_1+2 \beta_0 \dot{x}_1+\omega_0^2 x_1 -\gamma \dot{x}_2-g^2 x_2 =K_0 \mathrm{exp}(-\mathrm{i} \omega_{ex} t),
\\ &......
\\ &\ddot{x}_{j}+2 \beta_0 \dot{x}_{j}+\omega_0^2 x_{j}
\\ & -\gamma \dot{x}_{j-1}-g^2 x_{j-1}-\gamma \dot{x}_{j+1}-g^2 x_{j+1} =K_0 \mathrm{exp}(-\mathrm{i} \omega_{ex} t),
\\ & (\mathrm{for}\ 2 \leq j \leq n-1 )
\\ &......
\\ & \ddot{x}_{n}+2 \beta_0 \dot{x}_{n}+\omega_0^2 x_{n-1}-\gamma \dot{x}_{n-1} - g^2 x_{n-1}  =K_0 \mathrm{exp}(-\mathrm{i} \omega_{ex} t).
\end{aligned}
\label{eq:basic01}
\end{equation}
Here, $x_j(t)$, $\dot{x}_j(t)$ and $\ddot{x}_j(t)$ are the displacement relative to equilibrium position, velocity and acceleration of $j$-th oscillator, respectively, $K_0=-e E_0/m_e$, and $e$ is the charge of electron.

%~\\ \indent
% \subsection{Sca}
Firstly, we deal with the scattering properties.
Assume that $x_j(t)=A_j \mathrm{exp}(\alpha t)$ ($j=1,\ 2,\ ...,\ n$) are the solutions of Eq. (\ref{eq:basic01}).
Define $B=-\gamma \alpha - g^2$, $C=\alpha^2+2\beta_0 \alpha +\omega_0^2$, and
\begin{equation}
D=
\begin{pmatrix}
C & B &   &      &   &    &     \\
B & C & B &      &   &    &     \\
  & B & C &      &   &    &     \\
  &   &   &\ddots&   &    &     \\
  &   &   &      & C & B    \\
  &   &   &      & B & C    & B  \\
  &   &   &      &   & B    & C
\end{pmatrix}
_{n\times n},
\label{eq:matrixD}
\end{equation}
where $D_{j_1 j_2}=C$ for $j_1=j_2$, $D_{j_1 j_2}=B$ for $|j_1-j_2|=1$, and $D_{j_1 j_2}=0$ for other cases.
Also define:
\begin{equation}
\begin{aligned}
X=(A_1,\ ...,\ A_j, \ ..., \ A_n)^T,
\\ K=K_0(1,\ ...,\ 1, \ ..., \ 1)^T.
\end{aligned}
\label{eq:matrixXK}
\end{equation}
After substituting $x_j(t)$ into Eq. (\ref{eq:basic01}), the amplitudes should satisfy the following equation:
\begin{equation}
DX=K,
\label{eq:DXK1}
\end{equation}
where the solution of $\alpha$ is $\alpha=-\mathrm{i} \omega_{ex}$.
The solutions of the amplitudes are:
\begin{equation}
X=D^{-1}K,
\label{eq:DXK2}
\end{equation}
where $D^{-1}$ is the inverse of matrix $D$.
The total far field emission of the electric field is proportional to the accelerations of the oscillators \cite{Griffiths,PL0,PL01}:
\begin{equation}
\begin{aligned}
\ddot{x}(t)=\sum_{j=1}^n \ddot{x}_j(t)&= (-\mathrm{i}\omega_{ex})^2 \left ( \sum_{j=1}^n A_j \right ) \mathrm{exp}({-\mathrm{i}\omega_{ex}t})
\\ & =-\omega_{ex}^2 A^{(n)}\ \mathrm{exp}({-\mathrm{i}\omega_{ex}t}).
\end{aligned}
\end{equation}
Actually, $A^{(n)}$ satisfy:
\begin{equation}
A^{(n)}=\sum_{j=1}^n A_j  = \sum_{j=1}^n X_{j}=K_0\sum_{j_1=1}^n\sum_{j_2=1}^n D_{j_1 j_2}^{-1} \ .
\end{equation}
We give out the solutions of the amplitudes $A^{(n)}$ for $n=1-\-6$ as examples:
\begin{equation}
\begin{aligned}
   &A^{(1)}=\frac{1}{C}K_0,\ \ \ \ \ \ \ \ \ \ \ \ A^{(2)}=\frac{2 }{B+C} K_0,
\\ &A^{(3)}= \frac{4B-3C}{2B^2-C^2} K_0,\ \ A^{(4)}= \frac{2(B-2C)}{B^2-BC-C^2} K_0,
\\ &A^{(5)}= \frac{B^2+8BC-5C^2}{3B^2C-C^3} K_0,
\\ &A^{(6)}= \frac{2(2B^2+2BC-3C^2)}{B^3+2B^2C-BC^2-C^3} K_0\ .
\end{aligned}
\end{equation}
Define $A^{'(n)}=-\omega_{ex}^2 A^{(n)}$, and employ the Fourier transform in Ref. [\mycite{PL0,PL01}], thus obtaining the elastic emission spectrum:
\begin{equation}
I_{ela}(\omega)=\left |A^{'(n)} \right|^2\sqrt{2\pi}\delta(\omega-\omega_{ex}) .
\end{equation}
Therefore, the white light scattering spectra should be given as:
\begin{equation}
I_{sca}(\omega)=I_{ela}(\omega_{ex} \to \omega)=\sqrt{2\pi}\left|A^{'(n)}(\omega_{ex} \to \omega)\right|^2 .
\label{eq:Isca}
\end{equation}

%~\\ \indent
% \subsection{PL}
Secondly, we deal with the PL properties.
As our previous work demonstrates \cite{PL0}, PL term origins from the general solutions of the homogeneous linear equations:
\begin{equation}
DX=0.
\label{eq:DX0}
\end{equation}
The necessary and sufficient conditions for the existence of nontrivial solutions of Eq. (\ref{eq:DX0}) is that the determinant of $D$ is zero:
\begin{equation}
\mathrm{det}(D)=\frac{z_{+}^{n+1}-z_{-}^{n+1}}{z_{+}-z_{-}}=0,
\label{eq:detD}
\end{equation}
where $z_{\pm}=\frac{1}{2}(C\pm\sqrt{C^2-4B^2})$. Eq. (\ref{eq:detD}) determines the solutions of $\alpha$.
Obviously, there are $2n$ solutions for Eq. (\ref{eq:detD}). We can rewrite the solutions as $\alpha_{k\pm}=-\beta_k \pm \mathrm{i} \omega_k$, $k=1,\ 2,\ ...,\ n$. Due to the large difference from the excitation frequency, we omit the solutions of ``$\alpha_{k+}$'' as our previous work does \cite{PL0,PL01}. Therefore, the total solutions for Eq. (\ref{eq:basic01}) can be assumed and solved by:
\begin{equation}
\begin{aligned}
&x_j(t)=S_j \mathrm{exp}(-\mathrm{i}\omega_{ex} t) + \sum_{k=1}^{n} P_{jk} \mathrm{exp}(\alpha_{k-} t),
\\& \mathrm{initial\ conditions:}
\\& x_j(0)=0,\ \dot{x}_j(0)=0,\ \ddot{x}_j(0)=K_0,\ ...,\ \frac{\mathrm{d}^n x_j(0)}{\mathrm{d} t^n},
\\& \mathrm{for}\ j=1,\ 2,\ ...,\ n.
\end{aligned}
\end{equation}
Here, $S_j$ is the amplitude of the elastic term (scattering) of the $j$-th oscillator; $P_{jk}$ is the amplitude of the $k$-th inelastic term (PL) of the $j$-th oscillator.
After considering the PL term for the solutions, the total far field emission of the electric field can be written as:
\begin{equation}
\begin{aligned}
\ddot{x}(t)=\sum_{j=1}^{n}\ddot{x}_j(t)=&(-\mathrm{i}\omega_{ex})^2 \sum_{j=1}^{n}S_j \mathrm{exp}(-\mathrm{i}\omega_{ex} t)
\\+ & \sum_{j=1}^{n} \sum_{k=1}^{n} \alpha_{k-}^{2} P_{jk} \mathrm{exp}(\alpha_{k-} t).
\end{aligned}
\end{equation}
Define $P_{k}'=\sum_{j=1}^{n}\alpha_{k-}^{2} P_{jk}$, employing Fourier transform and Fermi-Dirac distribution \cite{PL0,PL01}, the total PL spectrum of the system can be written as:
\begin{equation}
\begin{aligned}
I_{PL}(\omega)=&\sum_{k=1}^{n} \left | P_{k}' \right |^2  \frac{1-\mathrm{exp}(-2 \beta_k t_0)}{2 \beta_k t_0} \frac{\beta_k}{(\omega-\omega_k)^2+\beta_k^2}
\\ & \times \frac{1}{1+\mathrm{exp}[(\hbar \omega -\hbar \omega_f)/k_B T]}.
\end{aligned}
\label{eq:IPL}
\end{equation}
Here, $t_0$ is the effective interaction time between the excitation light and the oscillators, $\hbar$ is the reduced Planck constant, $\omega_f$ is the so-called chemical potential and $k_B$ is Boltzmann's constant, and $T$ is the temperature.

For $n=1$, the solution degenerates into the one of an individual particle, which has been discussed in Ref. [\mycite{PL0}]. For $n=2$, the solutions degenerate into the dimer case, which has been discussed in Ref. [\mycite{PL01}].

~\\ \indent
%\section{\label{sec:Results}Results and Discussions}
Based on these formulas, we could analyze in detail to understand the scattering and PL properties more deeply. In the following statements, unit ``eV'' and unit ``Hz'' for $g$ satisfy the following relationship:
\begin{equation}
g[\mathrm{eV}]=\frac{\hbar}{e}g[\mathrm{Hz}].
\label{eq:eV}
\end{equation}
So do $\beta$, $\omega$ and $\gamma$.

%~\\ \indent
%\subsection{Scattering Results}
First of all, scattering properties are analyzed.

\begin{figure}[tb]
\includegraphics[width=0.48\textwidth]{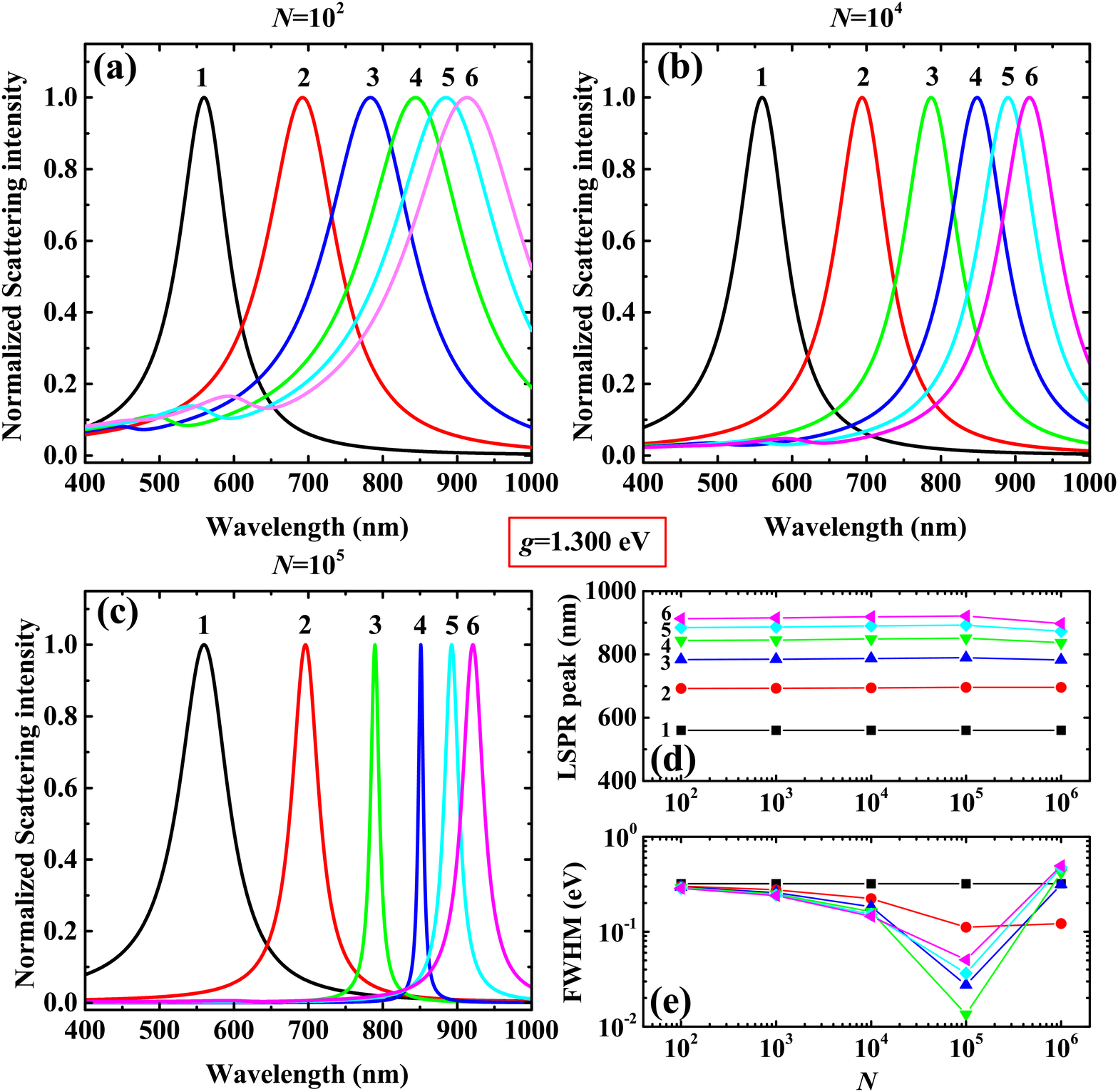}
\caption{\label{fig:changeN} (a)(b)(c) Normalized scattering spectra of the nanoparticle chains varying with effective free electron number $N$, i.e., $N=10^2$, $N=10^4$ and $N=10^5$, respectively, calculated from Eq. (\ref{eq:Isca}). (d) LSPR peak position of the chains as a function of $N$. (e) FWHM of the chains as a function of $N$. The spectra and curves are numerically labeled according to particle number $n$ except for (e). Black, red, blue, green, cyan and purple stand for $n$ equaling from 1 to 6, respectively. Here, the coupling strength is kept as $g=1.300$ eV.
}
\end{figure}

Fig. \ref{fig:changeN}a-c show the normalized scattering spectra of the chains with different particle number $n$ varying with effective free electrons number $N$. The coupling strength is $g=1.3$ eV. The primary LSPR peak red-shifts as $n$ increases. When $n \geq 3$, there exist other peaks at blue side of the primary peaks with small amplitudes for each case. Fig. \ref{fig:changeN}d shows the LSPR peak positions as a function of $N$ for different cases. It indicates that for small $n$, the peaks almost stay unchanged as the increase of $N$. However, for large $n$, the peak decrease at about $N=10^6$. Notice that in Fig. \ref{fig:changeN}, we keep $g$ unchanged, and Eq. (\ref{eq:g}) can be rewritten as $\gamma^3=g^4/\kappa$. When $N$ is small, $\gamma$ is much less then $g$, thus influencing the interaction parts little. However, when $N$ increases, $\kappa$ decreases, resulting in the increase of $\gamma$, which influences the interaction parts greatly if $\gamma$ is comparable with or even larger than $g$. Hence, the peak positions are influenced by $N$. Fig. \ref{fig:changeN}e shows FWHM of different cases as a function of $N$. For $n \geq 2$, FWHM decreases first and then increases as $N$ increases. The minimums of FWHM occur at about $N=10^5$, resulting in narrow shapes of the spectra as shown in Fig. \ref{fig:changeN}c.

\begin{figure}[tb]
\includegraphics[width=0.48\textwidth]{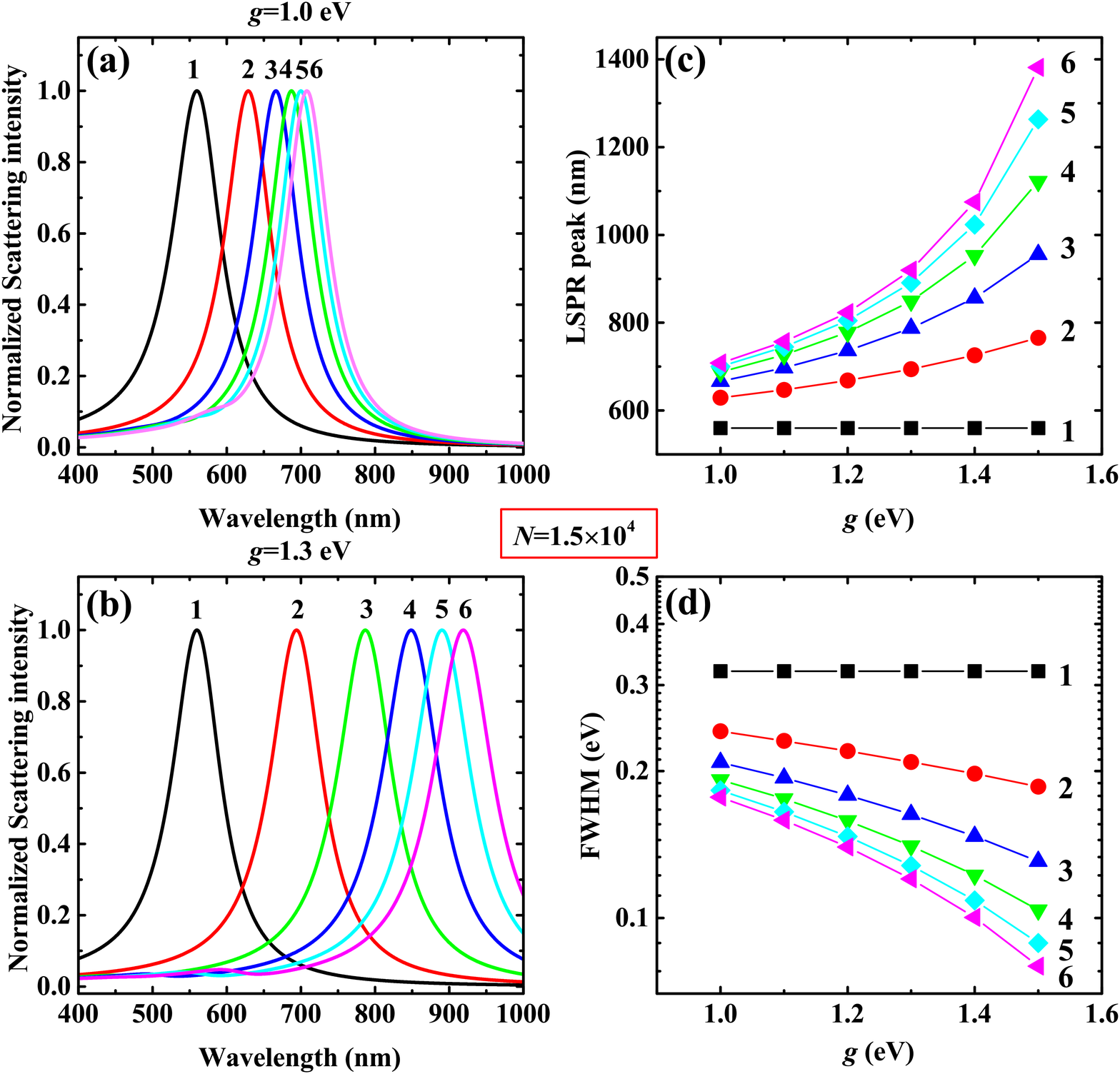}
\caption{\label{fig:changeg} (a)(b) Normalized scattering spectra of the nanoparticle chains varying with coupling strength $g$, i.e., $g=1.0$ eV and $g=1.3$ eV, respectively, calculated from Eq. (\ref{eq:Isca}). (c) LSPR peak position of the chains as a function of $g$. (d) FWHM of the chains as a function of $g$. The spectra and curves are numerically labeled according to particle number $n$. Here, the effective free electron number is kept as $N=1.5\times 10^4$.
}
\end{figure}

Fig. \ref{fig:changeg}a-b show the normalized scattering spectra of the chains with different $n$ varying with coupling strength $g$. Here, $N=1.5\times 10^4$. The same as Fig. \ref{fig:changeN}a-c, the primary LSPR peak red-shifts as $n$ increases. Obviously, the red-shift of $g=1.3$ eV is larger than the one of $g=1.0$ eV. In Fig. \ref{fig:changeg}c, LSPR wavelength increases as $n$ increases (red-shift), and it also increases as $g$ increases (red-shift). That is, the larger $g$ or/and $n$ is, the larger red-shift is. In Fig. \ref{fig:changeg}d, FWHM decreases as $n$ increases, and it also decreases as $g$ increases, resulting in narrower shapes of the spectra with larger $g$ or/and $n$ in unit of ``eV''.

\begin{figure}[tb]
\includegraphics[width=0.48\textwidth]{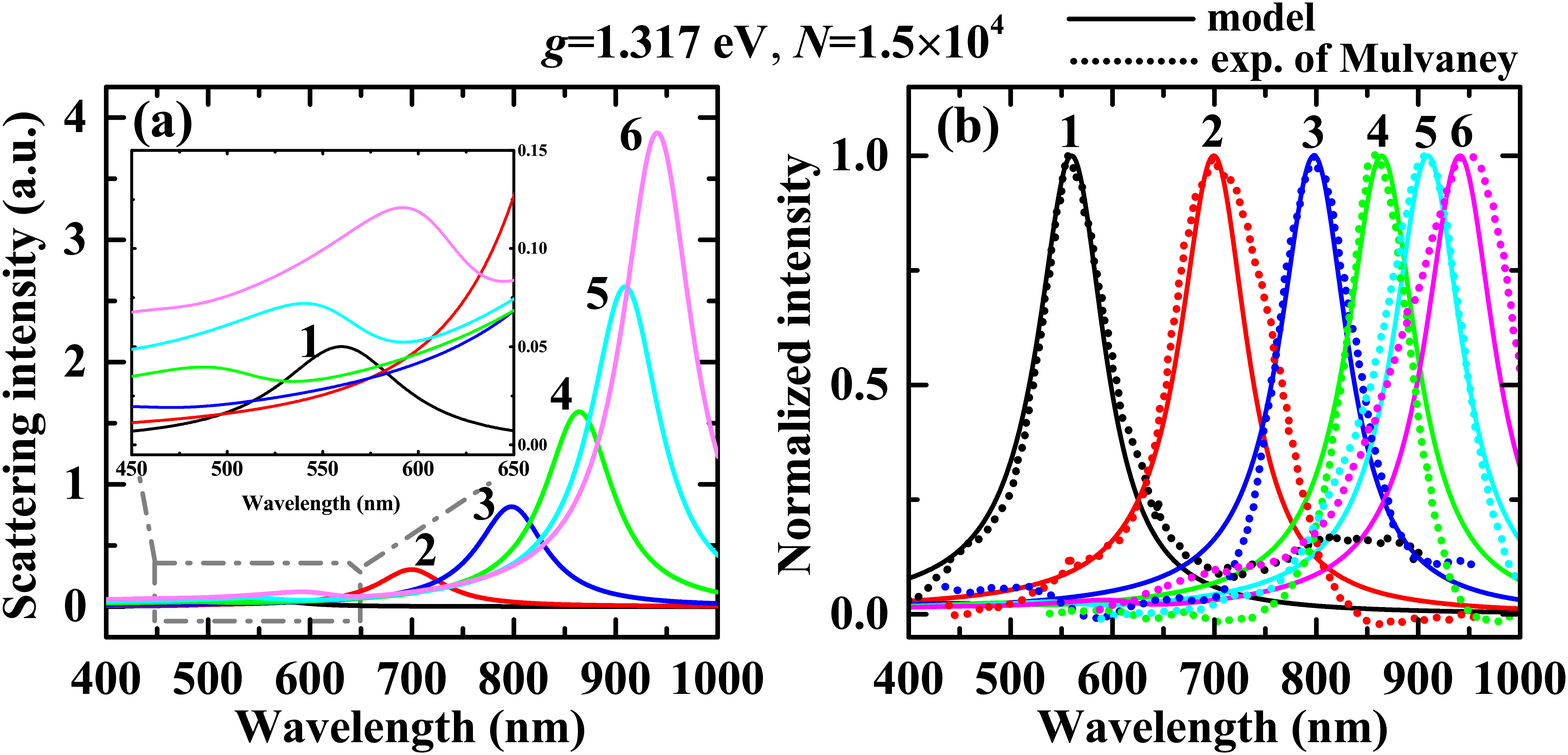}
\caption{\label{fig:contrast} (a) Scattering spectra of the nanoparticle chains calculated from Eq. (\ref{eq:Isca}). The inset of (a) is the zoom of the dashed box. (b) Normalized scattering spectra of the nanoparticle chains calculated from Eq. (\ref{eq:Isca}) (solid lines) and the experimental ones of P. Mulvaney \cite{couple4} (dot lines). The spectra are numerically labeled according to particle number $n$. Here, $N=1.5\times 10^4$, $\omega_0=2.204$ eV, $\beta_0=0.156$ eV and $g=1.317$ eV.
}
\end{figure}

In Fig. \ref{fig:contrast}, we show the comparison between the scattering spectra of these chains calculated from this model and the experimental ones from P. Mulvaney $et\ al$.\cite{couple4}. Here, the intrinsic frequency $\omega_0$ and damping coefficient $\beta_0$ of each identical nanoparticle are $\omega_0=2.204$ eV and $\beta_0=0.156$ eV, resulting in the resonant wavelength at 560 nm of an individual nanoparticle. The coupling strength and effective free electrons number are $g=1.317$ eV and $N=1.5\times10^4$.
Fig. \ref{fig:contrast}a shows the scattering spectra of these chains calculated from this model. As $n$ increases, the scattering intensity increases and meanwhile the peak red-shifts. The inset of Fig. \ref{fig:contrast}a illustrates the peaks with small values (for $n\geq 3$). The intensities of these small peaks increase and they also red-shift as $n$ increases. In Fig. \ref{fig:contrast}b, we compare the normalized scattering spectra of our model with the experiments of P. Mulvaney \cite{couple4}. In general, they agree well with each other. Notice that the FWHMs of $n=2$ and $n=6$ agree not very well. For $n=2$, the dimer case, the FWHM of our model is smaller than the one of the experiment. For $n=6$ the result reverses, and the peak of the model is a bit blue-shifted compared to the experiment. The main reason may be the nonuniformity of the samples. Of course, other reasons should be considered. For example, in our model, we omit the interaction parts of the non-neighboring particles, which indeed influences the spectra.

%~\\ \indent
%\subsection{PL Results}
The next, PL properties are analyzed.

\begin{figure}
\includegraphics[width=0.48\textwidth]{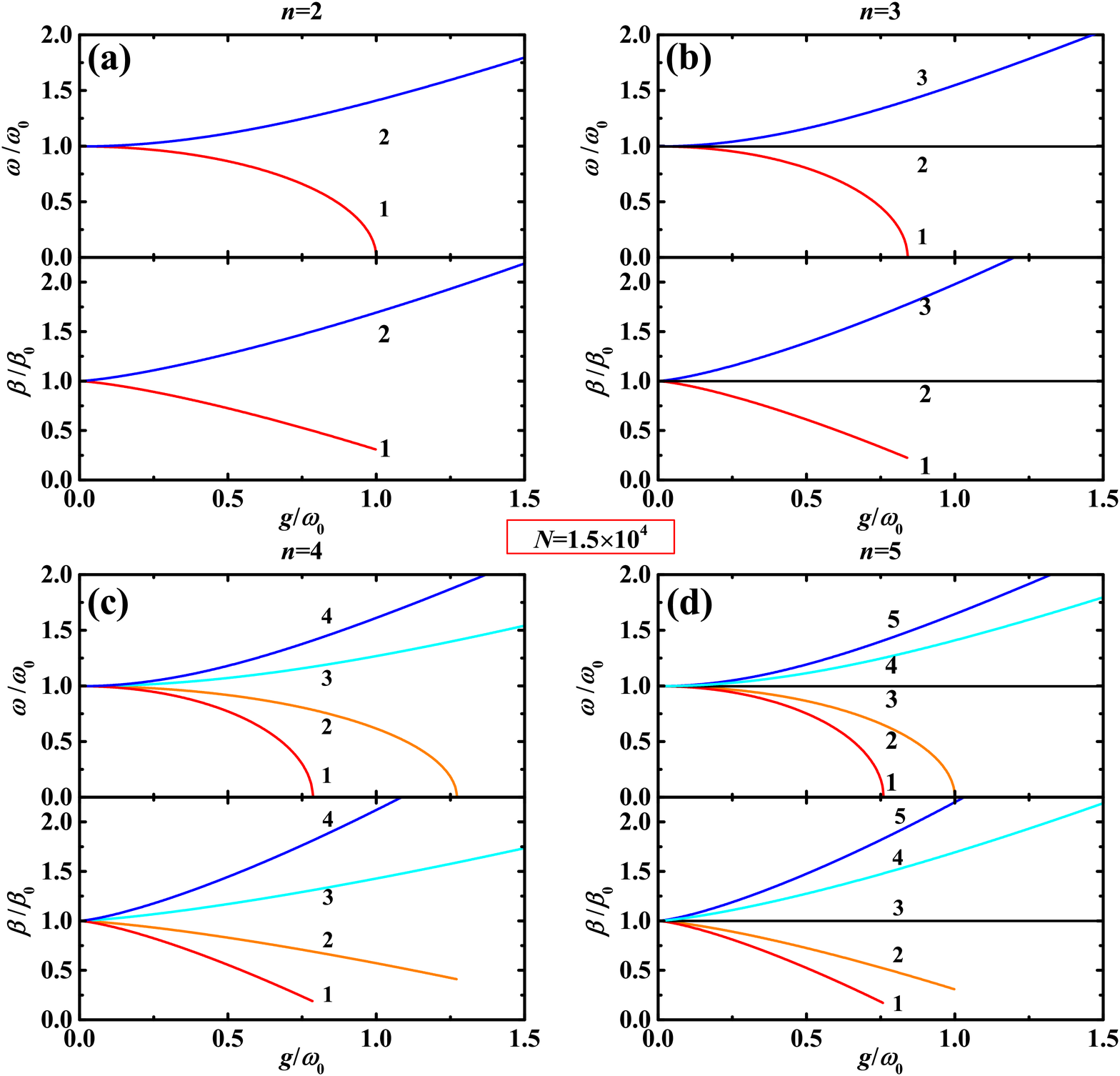}
\caption{\label{fig:PLchangeg} The new generated resonant frequencies ($\omega_j$) and damping coefficients ($\beta_j$) as a function of $g$ in the cases of $n=2$ (a), $n=3$ (b), $n=4$ (c) and $n=5$ (d), respectively, calculated from Eq. (\ref{eq:detD}). The curves are numerically labeled according to Mode $j$. Here, $N$, $\omega_0$ and $\beta_0$ are the same as the ones in Fig. \ref{fig:contrast}.
}
\end{figure}

The new generated PL modes are shown in Fig. \ref{fig:PLchangeg}, taking $n=2-\-5$ as examples. For a given particle number $n$, there are $n$ modes of PL. Particularly, when $n$ is even, the modes spit, half ($n/2$) of which blue shift, called the blue branches, the other half ($n/2$) of which red shift, called the red branches. When $n$ is odd, however, there exists one mode with frequency and damping unchanged and equaling to $\omega_0$ and $\beta_0$, respectively, and the rest modes split the same as even case. More special, Mode 1 and Mode 2 of $n=2$ behave the same as Mode 2 and Mode 4 of $n=5$. There exist cut-off coupling strengthes $g_{cut}$ for the red branches, at which the frequency decreases to zero. For Mode 1 of all the cases, $g_{cut}$ decreases as $n$ increases.
Notice that in Eq. (\ref{eq:IPL}), the shape of each mode is a Lorentzian curve with FWHM $2\beta_j$. Hence, the increase (decrease) of $\beta_j$ in Fig. \ref{fig:PLchangeg} indicates the increase (decrease) of FWHM of Mode $j$.

\begin{figure}
\includegraphics[width=0.48\textwidth]{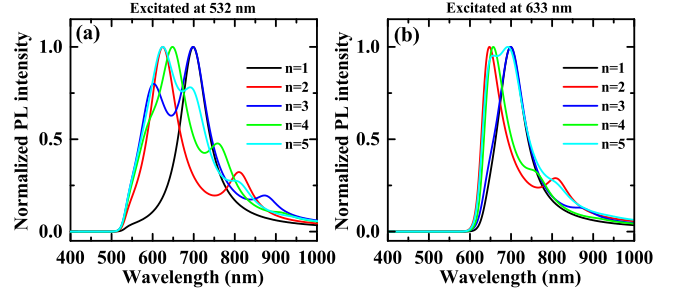}
\caption{\label{fig:IPL} Normalized PL spectra of nanoparticle chains calculated from Eq. (\ref{eq:IPL}), excited at 532 nm (a) and 633 nm (b), respectively. Black, red, blue, green and cyan stand for $n$ equaling from 1 to 5, respectively. Here, $N=1.5\times 10^4$, $g=0.9$ eV, $\omega_0=1.774$ eV and $\beta_0=0.101$ eV.
}
\end{figure}

In Fig. \ref{fig:IPL}, PL spectra calculated from Eq. (\ref{eq:IPL}) are shown for two different excitation wavelengths, i.e., 532 nm and 633 nm. In Fig. \ref{fig:IPL}a, it reveals the number of modes as 1, 2, 3, 4 and 3 for $n=$1, 2, 3, 4 and 5, respectively. There is only 3 modes in the spectrum of $n=5$, because one mode's wavelength is larger than 1000 nm so that it is not shown, the other mode's wavelength is smaller than 532 nm so that Fermi-Dirac distribution makes it disappear. In Fig. \ref{fig:IPL}b, due to the fact that some modes' wavelengths are smaller than 633 nm, they disappear. The numbers of the modes illustrated here are 1, 2, 2, 3 and 3 for $n=$ 1, 2, 3, 4 and 5, respectively. Notice that, for $n=2$ and $n=5$, there are two modes behaving similarly at around 624 nm and 811 nm, respectively. This phenomenon has been revealed in Fig. \ref{fig:PLchangeg}.

%~\\ \indent
%\subsection{before Conclusion}
Furthermore, we should emphasize that, in this work, the excitation electric field felt by these particles is treated as the same, i.e., the amplitudes and phases are the same. This approximation is reasonable when dealing with small $n$, e.g., $n\leq 6$. In Mulvaney's experiments \cite{couple4}, the diameter of one particle is 64 nm with gap of 1 nm, resulting in the length of the chain for $n=6$ of about 390 nm (omitting the bend of the chain). For larger $n$, the length of the chain would be so large that it is larger than the excitation wavelength. In experiments, usually, the light is focused through the lens and the size of the facula is near the excitation wavelength due to the diffraction limit. In other words, a large $n$ breaks the identity of the right side in Eq. (\ref{eq:basic01}), i.e., the electric field felt by the particles is not the same and it depends on the position of the facula. Therefore, if employing our model to calculate the scattering and PL spectra of a long chain, one just needs to rewrite Eq. (\ref{eq:basic01}) according to the situation, and then derives the rest equations as this work does.

~\\ \indent
%\section*{\label{sec:Conclusion}Conclusions}
In summary, we develop a multimer coupling classic harmonic oscillator model and employ it to explain the white light scattering and PL spectra of metallic nanoparticle chains which are strongly coupled. The model is suitable for $n \geq 1$, and we take $n=1-\-6$ as examples in this work. Comparisons with experiments of Mulvaney' work illustrate the accuracy and practicability of this model. Moreover, the scattering and PL properties are analyzed in detail, which depend on particle number $n$, coupling strength $g$ and effective free electron number $N$. For scattering spectra, larger $n$ or/and larger $g$ result in larger red-shift of LSPR peak and in smaller FWHM of the peak; small $N$, e.g., $N < 10^5$, influences the spectra little, while large $N$, e.g., $N \geq 10^5$, influences the spectra greatly. For PL spectra, the modes split due to the coupling and the splitting increases with the increase of $g$. The amplitudes of these modes are dependent on the excitation wavelength.
This classic model is practical and accurate when dealing with the coupling of metallic nanostructures. Thereby, this work would be helpful to understanding optical properties more deeply and gives a unified treatment for scattering and PL properties of strongly coupled multimer system. It also useful for related applications utilizing strongly coupled system of nanophotonics.

~\\ \indent% add an empty line
%\section*{\label{sec:Acknowldegment}Acknowledgment}
This work was supported by the Fundamental Research Funds for the Central Universities (Grant No. FRF-TP-20-075A1). Thanks to Professor Paul Mulvaney for the permission of using the data from his previous work.

%\section*{Disclosures}
The authors declare no conflicts of interest.

%\section*{Data availability}
The data that support the findings of this study are available from the corresponding author upon reasonable request.

\section*{\label{sec:Ref}References}
%\bibliography{viscosity_Cheng}
%\balance
\bibliography{chain_Cheng}

\end{document}